\documentclass[twocolumn,prb,superscriptaddress]{revtex4}
\usepackage{graphicx}
\usepackage{amsmath}

\begin{document}

\title{Electrical spin protection and manipulation via gate-locked spin-orbit fields}

\author{Florian Dettwiler}
\affiliation
{Department of Physics, University of Basel,
CH-4082, Basel, Switzerland}

\author{Jiyong Fu}
\thanks{Permanent address: Department of Physics, Qufu Normal University,
  Qufu, Shandong, 273165, China}
\affiliation
{Instituto de F\'isica de S\~ao Carlos, Universidade de S\~ao Paulo,
13560-970 S\~ao Carlos, SP, Brazil}

\author{Shawn Mack}
\thanks{Current address: Naval Research Laboratory, Washington, DC 20375, USA}
\affiliation
{Center for Spintronics and Quantum Computation, University of California, Santa Barbara,
California 93106, USA}

\author{Pirmin J. Weigele}
\affiliation
{Department of Physics, University of Basel,
CH-4082, Basel, Switzerland}

\author{J. Carlos Egues}
\affiliation
{Instituto de F\'isica de S\~ao Carlos, Universidade de S\~ao Paulo,
13560-970 S\~ao Carlos, SP, Brazil}

\author{David D. Awschalom}
\affiliation
{Center for Spintronics and Quantum Computation, University of California, Santa Barbara,
California 93106, USA}
\affiliation
{Institute for Molecular Engineering, University of Chicago, Chicago, IL  60637 USA}

\author{Dominik M. Zumb\"uhl}
\email{dominik.zumbuhl@unibas.ch}
\affiliation
{Department of Physics, University of Basel,
CH-4082, Basel, Switzerland}


\maketitle

\thispagestyle{plain}

\bfseries
The spin-orbit (SO) interaction couples electron spin and momentum via a relativistic, effective magnetic field. While conveniently facilitating coherent spin manipulation\cite{datta:1990,golovach:2006,nowack:2007} in semiconductors, the SO interaction also inherently causes spin relaxation\cite{golovach:2004,kroutvar:2004,amasha:2008}. A unique situation arises when the Rashba and Dresselhaus SO fields are  matched\cite{schliemann:2003,bernevig:2006,Duckheim:2009}, strongly protecting spins from relaxation, as recently demonstrated\cite{koralek:2009,walser:2012_1}. Quantum computation and spintronics devices such as the paradigmatic spin transistor\cite{datta:1990,koo:2009} could vastly benefit if such spin protection could be expanded from a single point into a broad range accessible with \emph{in-situ} gate-control, making possible tunable SO rotations under protection from relaxation. 

Here, we demonstrate broad, independent control of all relevant SO fields in GaAs quantum wells, allowing us to tune the Rashba and Dresselhaus SO fields while keeping both \emph{locked} to each other using gate voltages. Thus, we can electrically control and simultaneously protect the spin. Our experiments employ quantum interference corrections to electrical conductivity as a sensitive probe of SO coupling. Finally, we combine transport data with numerical SO simulations to precisely quantify all SO terms.

\begin{figure}[*t]\vspace{-1mm}
\includegraphics[width=8.7cm]{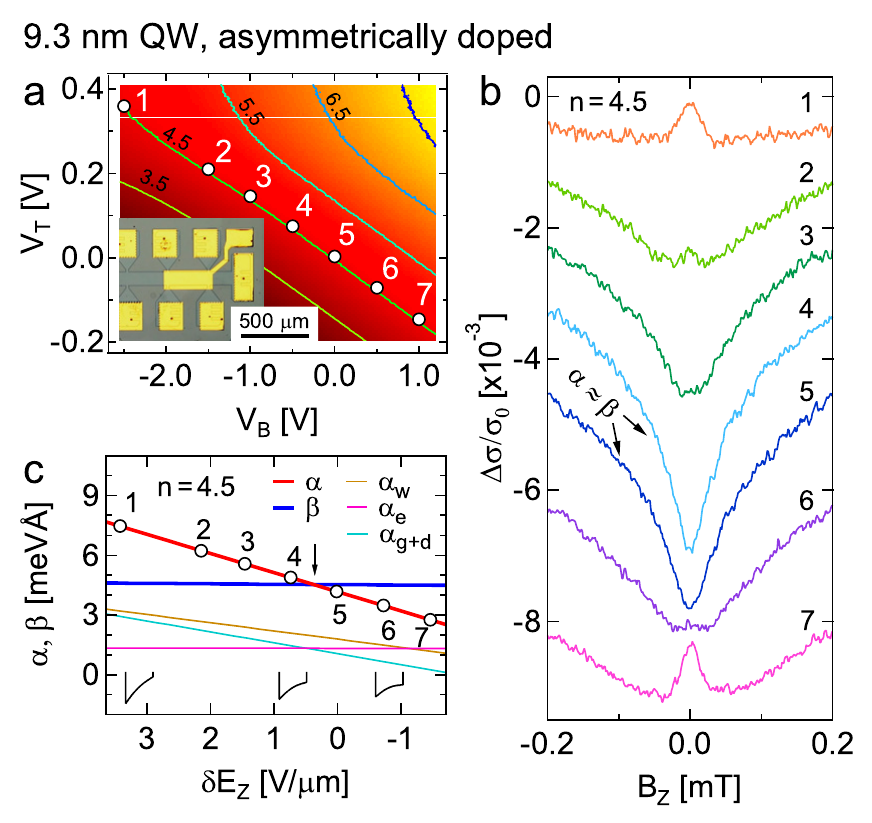}
\vspace{-4mm}
\caption{ {\bf Weak localization (WL) as an $\alpha=\beta$ detector; gate-control of Rashba $\alpha$ at constant density.} {\bf a} Measured charge density $n$ (color) versus top gate voltage $V_T$ and back gate voltage $V_B$ ($9.3\,$nm QW). Contours of constant density $3.5 - 7.5\cdot 10^{11}\,\mathrm{cm^{-2}}$ are shown. Inset: optical micrograph of typical Hall bar, with contacts (yellow), gate (center) and mesa (black lines). {\bf b} Normalized longitudinal conductivity $\Delta \sigma / \sigma_0=(\sigma(B_Z)-\sigma(0))/\sigma(0)$ versus $B_Z \bot$ 2DEG. Curves for gate configurations $1-7$ along constant $n=4.5\cdot 10^{11}\,\mathrm{cm^{-2}}$ are shown (offset vertically), also labeled in panels {\bf a} and {\bf c}. {\bf c} Simulated Rashba $\alpha$ and Dresselhaus $\beta$ coefficients (see text) against gate-induced field change $\delta E_Z$, shown for constant $n=4.5\cdot10^{11}\,\mathrm{cm^{-2}}$. The $\delta E_Z$ axis -- decreasing from left to right -- corresponds exactly to the $V_B$ abscissa of panel {\bf a} for a covarying $V_T$ such that $n=4.5\cdot10^{11}\,\mathrm{cm^{-2}}$ constant. Sketches of the QW potential at 1, 4 and 6 illustrate the change of $\alpha$ with $\delta E_Z$. Note that $\alpha(\delta E_Z=0)\neq 0$ since the external E-field (see SOM) is not zero at $\delta E_Z=0$.}
\label{fig:1}\vspace{-6mm}
\end{figure}

\normalfont
Spin-orbit coupling in a GaAs 2D electron gas (2DEG) has two dominant contributions: the Rashba\cite{rashba:1984} and the Dresselhaus\cite{dresselhaus:1955} effects, arising from breaking of structural and crystal inversion symmetry, respectively. The Rashba coefficient $\alpha$ is proportional to the electric field in the quantum well (QW) and can be tuned with the doping profile\cite{koralek:2009} as well as \emph{in-situ} using gate voltages\cite{engels:1997,nitta:1997}. We achieve independent control of $\alpha$ and carrier density $n$ by using both top and back gates\cite{papadakis:1999, grundler:2000}, at voltages $V_T$ and $V_B$, respectively. To modify $\alpha$ at constant density, a change of $V_T$ can be compensated by an appropriate change of $V_B$ (see Fig.\,\ref{fig:1}a), controlling $\alpha$ via a change of the gate-induced average electric field $\delta E_Z$, where $z\bot$2DEG. Note that $\alpha$ can be continuously tuned to change sign.

\begin{figure}\vspace{-3mm}
\includegraphics[width=8.7cm]{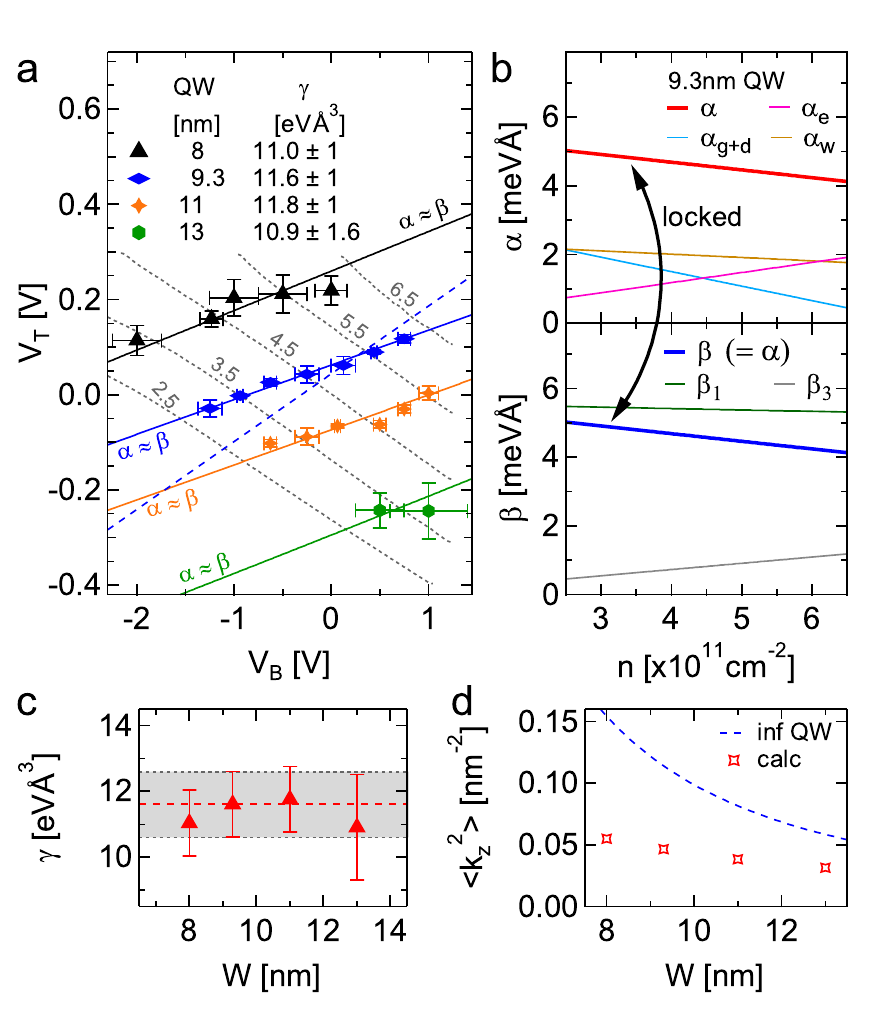}
\vspace{-8mm}
\caption{{\bf Tuning and locking $\alpha=\beta$.}
{\bf a} The markers indicate $\alpha\approx\beta$ for four different QW widths (asymmetric doping) and various densities (gray contours of constant $n$, labeled in units of $10^{11}\,\mathrm{cm^{-2}}$) in the $V_T$ and $V_B$ plane. Error bars result from the finite number of MC traces in the $(V_B,V_T)$-space. Theory fits (solid lines) are shown for each QW, with $\gamma$ as the only fit parameter (inset table, error bars dominated by systematic error, see below). The dashed blue line indicates the slope of constant $\alpha=\beta_1$, neglecting $\beta_3$, which is inconsistent with the data. {\bf b} Simulation of locked $\alpha=\beta$ versus density $n$ along solid blue line from {\bf a}, showing the various SO contributions (see text). {\bf c} Values of $\gamma$ from fits for each well width $W$. Red dashed line is the average $\gamma=11.6\pm1\,\mathrm{eV\mathring{A}^3}$ (excluding $W=13\,\mathrm{nm}$ due to its larger error), gray the $\sim9\%$ error, stemming mostly from the systematic uncertainty in the input parameters of the simulations (see methods).  {\bf d} $\langle k_z^2\rangle$ as a function of QW width $W$ for realistic (markers) and infinite (blue) potential.
}\label{fig:2}\vspace{-5mm}
\end{figure}

The Dresselhaus term\cite{dresselhaus:1955}, on the other hand, can be decomposed into sin/cos functions\cite{iordanskii:1994} of $\theta$ (1st harmonic) and $3\theta$ (3rd harmonic), with polar coordinate $\theta$ of the momentum $\textbf{k}$ in the 2DEG plane. The first harmonic has a coefficient $\beta=\beta_1-\beta_3$ comprised of $\beta_1=\gamma\langle k_z^2\rangle$ and $\beta_3=\gamma k_F^2/4$, where $k_F=\sqrt{2\pi n}$ is the Fermi wave number, $\gamma>0$ is the bulk Dresselhaus parameter and $\langle k_z^2\rangle$ depends on the width W of the well\cite{luo:1990}, with $\langle k_z^2\rangle=(\pi/W)^2$ for infinite barriers. In addition, $\beta_3$ is the coefficient of the 3rd harmonic Dresselhaus term -- hereafter cubic term -- and depends on density $n$, thus enabling gate-control. The cubic term breaks the angular symmetry of the other terms and is relatively small compared to $\beta_1$ since $\langle k_z^2 \rangle \gg k_F^2/4$, except for the largest densities used. To control $\beta_1$, we studied QWs of widths $W=8, 9.3, 11$ and $13\,$nm, yielding wave functions spreading over the full well width and $\langle k_z^2\rangle<(\pi/W)^2$, see Fig.\,\ref{fig:2}d, due to wave function penetration into the finite barriers. Over this range, $\beta_1$ changes by a factor $2$ and is essentially independent of gate voltage.

Thus, we can set $\beta_1$ in discrete steps via QW width, and we achieve independent, continuous control of $\alpha$ and $\beta_3$ \emph{in-situ} using top and back gates. This wide tunability makes it possible to change $\alpha$ while \emph{locking} $\beta$ to match $\alpha$ via gate-control of $\beta_3$, thus strongly protecting spins from relaxation. In previous experiments, the Rashba and Dresselhaus fields were matched only at isolated points\cite{koralek:2009,walser:2012_1,kohda:2012}, not over a broad range. Finally, the broad gate-tunability gives us access to various SO regimes, such as the Dresselhaus (Rashba) regime $\beta\gg\alpha$ ($\alpha\gg\beta$), and allows a detailed characterization of the cubic term, which limits the spin lifetime even at $\alpha=\beta$.

Weak antilocalization (WAL) is a well established signature of SO coupling in magnetoconductance (MC) $\sigma(B_Z)$ \cite{bergmannreview, AAreview, iordanskii:1994, pikus:1995, knap:1996, miller:2003}. For the $|\alpha|=\beta\gg\beta_3$ regime, the resulting internal SO magnetic field $\textbf{B}_{\mathrm{int}}(\textbf{k})$ (see SOM) is uniaxial and vanishes for $\textbf{k}$ along one direction in the 2DEG plane ($\hat{x}_{-}$ for $\beta=+\alpha$, or $\hat{x}_{+}(\bot \hat{x}_{-})$ for $\beta=-\alpha$). Therefore, WAL (maximal $\sigma(B_Z=0)$) is suppressed and the effectively spin-less situation exhibiting weak localization (WL) (minimal $\sigma(B_Z=0)$) is restored \cite{pikus:1995,schliemann:2003,bernevig:2006,kohda:2012}. We introduce the SO lengths $\lambda_{\pm}=\hbar^2/(2m^*\left|\alpha\pm\beta\right|)$ (with Planck constant $\hbar$ and effective mass $m^*$) as the \emph{ballistic} length to be travelled along $\hat{x}_{\pm}$, respectively, for a spin rotation of one radian. For $\beta=+\alpha$, $\lambda_-$ diverges (no precessions) while $\lambda_+$ is finite, and vice versa for $\beta=-\alpha$. We note that a closed-form theory of MC including all relevant SO terms is not currently available\cite{iordanskii:1994,pikus:1995,knap:1996,miller:2003}, making extraction of the SO parameters from MC difficult.

Figure\,\ref{fig:1}b displays the MC of the $9.3\,$nm QW for configurations labeled $1-7$, showing a transition from WAL ($1\,\&\,2$) to WL ($4\,\&\,5$) back to WAL ($7$) upon monotonically changing $\alpha$ on a contour of constant density, see Fig.\,\ref{fig:1}a. Selecting the most pronounced WL curve allows us to determine the symmetry point $\alpha\approx\beta$. This scheme is repeated for a number of densities, yielding the symmetry point as a function of $n$. Thus, Rashba and Dresselhaus fields can be locked with gate voltages over a wide range (see Fig.\,\ref{fig:2}a, blue markers), varying $n$ by a factor of $2$. Numerical simulations\cite{calsaverini:2008} (see methods) can accurately calculate $\alpha$ and $\langle k_z^2 \rangle$. This leaves only one fit parameter: $\gamma$, the bulk Dresselhaus coefficient, which can now be extracted from fits to the density dependence of the symmetry point, see solid blue line in Fig.\,\ref{fig:2}a, giving excellent agreement. Further, the WL dip -- often used to determine phase coherence -- sensitively depends on the SO coupling (e.g. curves $3-6$ in Fig.\,\ref{fig:1}b), even before WAL appears. Negligence of SO coupling could thus lead to spurious or saturating coherence times.

Next, we change the Dresselhaus coefficient $\beta_1$ by varying the QW width and repeating the above procedure, summarized in Fig.\,\ref{fig:2}a. For wider QWs, a smaller $\beta_1$ results, shifting the symmetry point towards smaller $\alpha$, i.e. towards the lower right in Fig.\,\ref{fig:2}a. As seen, locking of $\alpha=\beta$ is achieved in all present wavers. Further, reduced $\beta_1$ makes possible the Rashba regime $\alpha\gg\beta$, particularly in the upper left of Fig.\,\ref{fig:2}a where $\alpha$ is large. Again performing fits over the density dependence of the symmetry point for each QW width, we obtain very good agreement, see Fig.\,\ref{fig:2}a, and extract $\gamma=11.6\pm1\,\mathrm{eV\mathring{A}^3}$ consistently for all QWs (Fig.\,\ref{fig:2}c). We emphasize that $\gamma$ is notoriously difficult to calculate and measure\cite{knap:1996,miller:2003}; the value reported here agrees well with recent studies\cite{krich:2007,walser:2012_1, walser:2012_2}.

The Rashba coefficient is composed of $\alpha=\alpha_{\rm g+d}+\alpha_{\rm w}+\alpha_{\rm e}$ in the simulation, with gate and doping term $\alpha_{\rm g+d}$, QW structure term $\alpha_{\rm w}$, and Hartree term $\alpha_{\rm e}$. Along a contour of constant density, mainly $\alpha_{\rm g+d}$ and $\alpha_{\rm w}$ are modified, while $\alpha_{\rm e}$ and $\beta$ remain constant, see Fig.\,\ref{fig:1}c. The density dependence for locked $\alpha=\beta$ (see Fig.\,\ref{fig:2}b)) shows that while $\beta_1$ is nearly constant, $\beta_3$ is linearly increasing with $n$, thus reducing $\beta=\beta_1-\beta_3$, and therefore forcing a corresponding reduction of $\alpha$ to keep $\alpha=\beta$ matched. The Hartree term $\alpha_{\rm e}$, however, increases for growing $n$. This is compensated by the other $\alpha$-terms to nicely match $\beta(n)$, see Fig.\,\ref{fig:2}b. Here, a change of $\sim25\%$ in SO strength ($\sim4$ to $\sim5\,\mathrm{meV \AA}$) corresponds to a gate-controlled spin rotation of $\pi/2$ over a precession length of $5\,\mathrm{\mu m}$, which is clearly shorter than the spin diffusion length over the whole range, thus enjoying protection from spin relaxation.

\begin{figure}\vspace{-1mm}
\includegraphics[width=8.7cm]{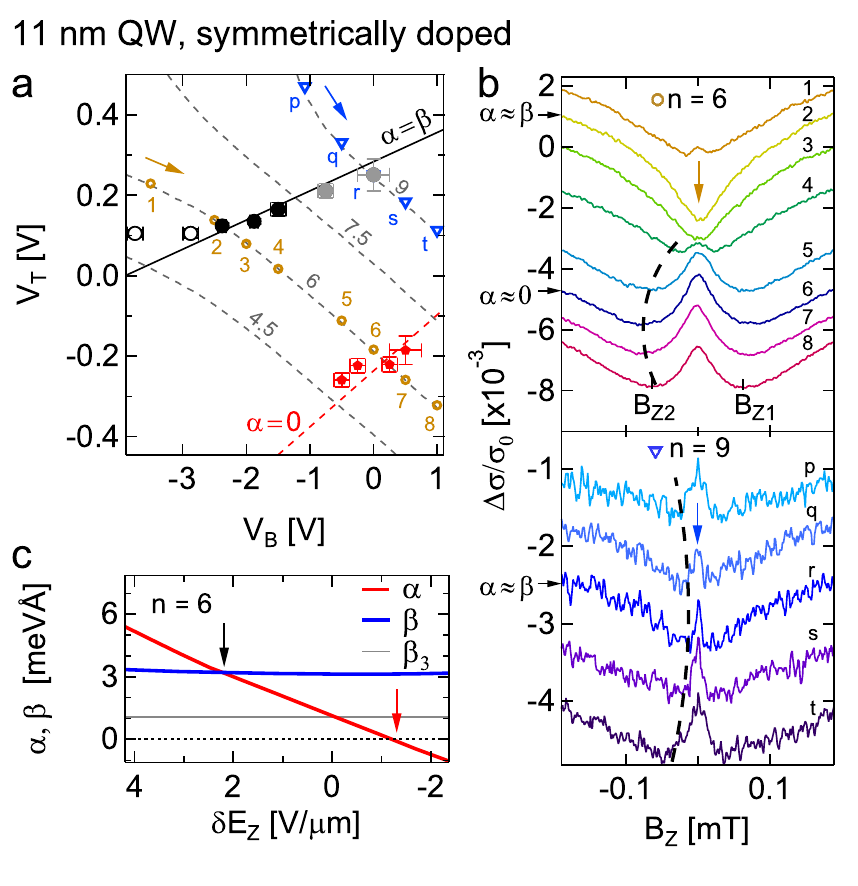}
\vspace{-8mm}
\caption{{\bf The Dresselhaus and the cubic regime.} {\bf a} Locked regime $\alpha\approx\beta$ (black/grey symbols) and Dresselhaus regime $\alpha\approx0$ (red symbols) from the broadest WAL minima (maximal $B_{\rm SO}$) in the $V_T$ and $V_B$ plane for a more symmetrically doped $11$\,nm QW. The solid black line displays the $\alpha=\beta$ simulation, while the dashed red line marks the simulated $\alpha=0$ contour. Open black markers (leftmost $V_B$) are entering the non-linear gate regime, causing a slight deviation from theory, which assumes linear gate action. The rightmost $V_B$ points (gray) are obtained from the minimal $B_{\rm SO}$ in presence of WAL. {\bf b} MC sequences at $n=6\cdot10^{11}\,\mathrm{cm^{-2}}$ (upper panel) and $n=9\cdot10^{11}\,\mathrm{cm^{-2}}$ (lower panel), shifted vertically for clarity. Each brown/blue marker in {\bf a} corresponds to a trace in {\bf b}, as labeled by numerals/letters. $B_{\rm SO}$ is indicated as a guide for the eye by black dashed curves for negative $B_Z$. {\bf upper panel} $B_{\rm SO}$ increases and peaks (indicating $\alpha=0$) before decreasing again. {\bf lower panel} broken spin symmetry regime: WAL is no longer suppressed here due to symmetry breaking from the cubic term at large $n$. Still, $\alpha\approx\beta$ can be identified with the narrowest WAL peak.  {\bf c} Simulation of $\alpha$ and $\beta$ along $n=6\cdot10^{11}\,\mathrm{cm^{-2}}$. $\alpha$ traverses both $\beta$ (black arrow) and for smaller $\delta E_Z$ also zero (red arrow).
}\label{fig:3}\vspace{-5mm}
\end{figure}

We now show that $\alpha$ can be tuned through $\beta$ \emph{and through zero} in a more symmetrically doped wafer, opening the Dresselhaus regime $\beta\gg\alpha$. We introduce the magnetic field $B_{\rm SO}$ where the MC exhibits minima at $B_{\rm Z1}\approx-B_{\rm Z2}$. Beyond the WAL-WL-WAL transition (Fig.\,\ref{fig:3}b upper panel), $B_{\rm SO}$ is seen to peak and decrease again (dashed curve). The gate voltages with maximal $B_{\rm SO}$ are added to Fig.\,\ref{fig:3}a for several densities (red markers). We surmise that these points mark $\alpha\approx0$: $B_{\rm SO}$ signifies the crossover between WL/WAL-like MC, thus defining an empirical measure for the effects of SO coupling (larger $B_{\rm SO}$, stronger effects). For $\alpha=0$, the full effect of $\beta$ on MC becomes apparent without cancellation from $\alpha$, giving a maximal $B_{\rm SO}$. Indeed, the simulated $\alpha=0$ curve (dashed red line in Fig.\,\ref{fig:3}a) cuts through the experimental points, also reflected in Fig.\,\ref{fig:3}c by a good match with the simulated $\alpha=0$ crossing point (red arrow).

\begin{figure}[!h]
\vspace{-2mm}
\includegraphics[width=8.7cm]{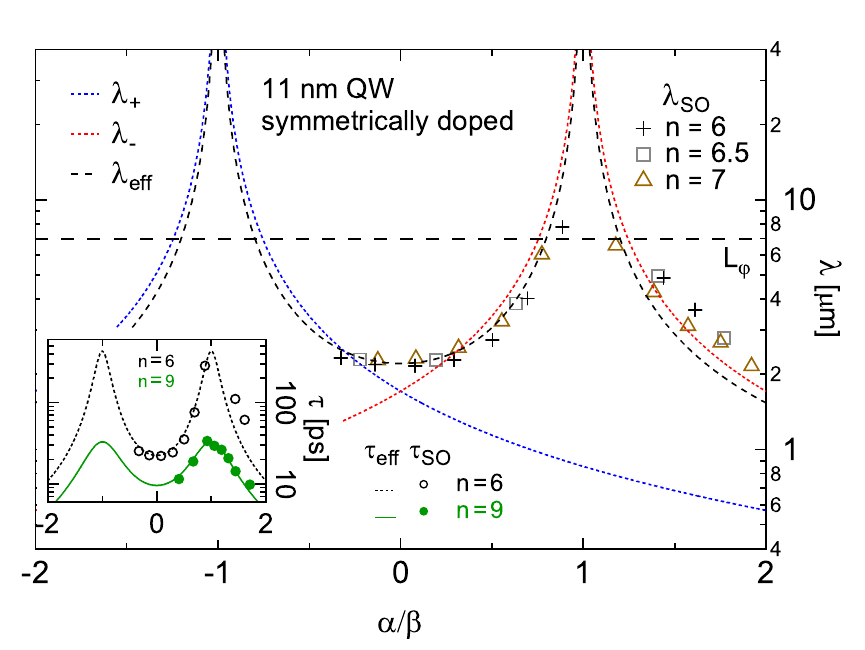}
\vspace{-7mm}
\caption{{\bf Experimental and theoretical SO-lengths and SO-times.} Experimental $\lambda_{\rm SO}=\sqrt{\hbar/2eB_{\rm SO}}$ (markers, densities as labeled, in units of $10^{11}\,\mathrm{cm^{-2}}$) as a function of the dimensionless ratio $\alpha/\beta$ (from SO simulation). The ballistic $\lambda_{\pm}$ (blue/red dashes) and effective $\lambda_{\mathrm{eff}}$ (black dashed curve) are only weakly $n$-dependent (small $\beta_3$) when plotted against $\alpha/\beta$. Thus, curves for only one density ($n=6\cdot10^{11}\,\mathrm{cm^{-2}}$) are shown. The experimental uncertainty on $\lambda_{\rm SO}$ is captured by the spread given by the three slightly different densities. The coherence length $L_\varphi\approx7\,\mathrm{\mu m}$ is added for illustration (obtained from WL curves), setting the visibility of SO effects in MC and thus the width of the WAL-WL-WAL transition. Inset: experimental spin relaxation time $\tau_{\rm SO}=\lambda^2_{\rm SO}/(2D)$ (circles) as a function of $\alpha/\beta$ for two densities as indicated. Theory curves $\tau_{\mathrm{eff}}$ (dashed) now include the symmetry breaking third harmonic term, preventing divergence at $\alpha/\beta=1$, while $\lambda_{\rm eff}$ (main panel) does not.}
\label{fig:4}\vspace{-5mm}
\end{figure}

For a comparison of experiment and simulation, we convert the empirical $B_{\rm SO}$ to a magnetic length $\lambda_{\rm SO}=\sqrt{\hbar/2eB_{\rm SO}}$, where $e>0$ is the electron charge and the factor of two accounts for time-reversed pairs of closed trajectories. Fig.\,\ref{fig:4} shows the theoretical spin diffusion length $\lambda_{\mathrm{eff}}$ (see methods) and the ballistic $\lambda_\pm$, together with the experimental $\lambda_{\rm SO}$, all agreeing remarkably well. First, we note that the ballistic $\lambda_\pm$ and the diffusive $\lambda_{\rm eff}$ (small $\beta_3$) are equivalent. The enhanced $\lambda_{\rm SO}$ around $\alpha/\beta=1$ corresponds to an increased spin relaxation time $\tau_{\rm SO}=\lambda_{\rm SO}^2/(2D)$. Second, $\max(\lambda_\pm)$ quantifies the deviation from the \emph{uniaxial} $\textbf{B}_{\mathrm{int}}(\textbf{k})$ at $\alpha=\beta$ and thus the extent to which spin rotations are not undone in a closed trajectory due to the non-Abelian nature of spin rotations around non-collinear axes. This leads to WAL, a finite $B_{\rm SO}$ and $\lambda_{\rm SO}\simeq \max(\lambda_\pm)$, as observed (see Fig.\,\ref{fig:4}). Unlike the corresponding time scales, the SO lengths are only weakly dependent on density and mobility when plotted against $\alpha/\beta$, allowing a comparison of various densities.

The cubic term causes spin relaxation even at $\alpha=\beta$ and becomes visible at large densities: WAL is present in all traces and through $\alpha=\beta$ (Fig.\,\ref{fig:3}b, lower panel), because $\textbf{B}_{\mathrm{int}}(\textbf{k})$ can no longer be made uniaxial, thus breaking spin symmetry and reviving WAL. A partial symmetry restoration is still apparent, where -- in contrast to the $\alpha=0$ case -- a \emph{minimal} $B_{\rm SO}$ is reached (dashed curves) consistent with $\alpha=\beta$ (grey markers Fig.\,\ref{fig:3}a at large $n$). We include the cubic $\beta_3$ in the spin relaxation time $\tau_{\mathrm{eff}}$ (see methods), shown in the inset of Fig.\,\ref{fig:4} for two densities, finding good agreement with the experimental $\tau_{\rm SO}=\lambda^2_{\rm SO}/(2D)$, where $D$ is the diffusion constant. Over the whole locked regime of Fig.\,\ref{fig:2}b, WAL is absent, and $\tau_{\rm SO}$ is enhanced between one and two orders of magnitude compared to $\alpha=0$. Finally, the coherence length $L_\varphi$ sets an upper limit for the visibility of SO effects: WAL is suppressed for $\lambda_{\rm eff}\gg L_\varphi$, setting the width of the WAL-WL-WAL transition (see SOM).

As an outlook, this work is laying the foundation for a new generation of experiments benefiting from unprecedented command over SO coupling in semiconductor nanostructures such quantum wires, quantum dots, and electron spin qubits. We note that analogous SO control is in principle also possible in other semiconductors. Finally, SO coupling is crucial beyond spintronics and quantum computation, e.g. for novel states of matter such as helical states, topological insulators, Majorana fermions and parafermions.

\vspace{6mm}
{\bf Materials and Methods}

{\bf GaAs QW materials.}
\small The QWs are grown on an n-doped substrate (for details see SOM) and fabricated into Hall bar structures (see inset, Fig.\,\ref{fig:1}a) using standard photolithographic methods. The 2DEG is contacted by thermally annealed GeAu/Pt Ohmic contacts, optimized for a low 2DEG contact resistance while maintaining high back gate tunability (low leakage currents) and avoiding short circuits to the back gate. On one segment of the Hall bar, a Ti/Au top gate with dimensions of $300$\,x\,$100$\,$\mu m^2$ was deposited. The average gate-induced E-field change in the QW is defined as $\delta E_Z=1/2\left(V_T/d_T-V_B/d_B\right)$, with effective distance $d_{T/B}$ from the QW to the top/back gate, extracted using a capacitor model, consistent with the full quantum description (see SOM). Contours of constant density follow $\delta V_T/d_T=-\delta V_B/d_B$. Deviations from linear behavior appear at most positive/negative gate voltages due to incipient gate leakage and hysteresis.

{\bf Low temperature electronic measurements.}
The experiments are performed in a dilution refrigerator with base temperature $20$\,mK. We have used a standard four-wire lock-in technique at $133\,$Hz and $100\,$nA current bias, chosen to avoid self-heating while maximising the signal. The density is determined with Hall measurements in the classical regime, whereas Shubnikov-de Haas oscillations were used to exclude occupation of the second subband, which is the case for all the data discussed. The WAL signature of MC is a small correction ($10^{-3}$) to total conductance. To achieve a satisfactory signal-to-noise ratio, longitudinal conductivity traces $\Delta\sigma/\sigma_0=(\sigma(B)-\sigma(0))/\sigma(0)$ were measured at least $10$ times and averaged.

{\bf Numerical Simulations.}
The simulations calculate the Rashba coefficient $\alpha$ and $\langle k_z^2\rangle$ based on the bulk semiconductor band parameters, the QW structure, the measured QW electron densities and the measured gate lever arms. We solve the Schr\"odinger and Poisson equations self consistently (``Hartree approximation''), obtain the self-consistent eigenfunctions, and then determine $\alpha$ via appropriate expectation values \cite{calsaverini:2008}. The Dresselhaus coefficient $\gamma$ is extracted from fits of the simulation to the experiment which detects the absence of WAL at $\alpha=\beta=\gamma(\langle k_z^2\rangle-k_F^2/4)$. Thus, given $\alpha$ and $\langle k_z^2 \rangle$ from the simulation and the measured $n=k_F^2/(2\pi)$, we obtain $\gamma=11.6\pm1\,\mathrm{eV\AA^3}$ consistently for all asymmetrically doped QWs. Taking into account the uncertainties of the band parameters, the experimental errors and a negligible uncertainty on $\langle k_z^2 \rangle$, an overall uncertainty of about $9-10\%$ or about $\pm 1\,\mathrm{eV\AA^3}$ on $\gamma$ results. About $1-2\%$ error originates from the experimental uncertainty of determining $\alpha=\beta$. The doping distribution (above/below QW) is not expected to influence $\gamma$, and hence we use the same $\gamma$ for the more symmetrically doped wafer. Fits to the $\alpha=\beta$ experimental points then determine how much charge effectively comes from upper rather than lower doping layers, fixing the last unknown parameter also for the more symmetrically doped QW (see SOM).

{\bf Spin-dephasing times and lengths.} In WL/WAL measurements, additional spin dephasing is introduced by the external magnetic field $B$ via the Aharonov-Bohm phase arising from the magnetic flux enclosed by the time reversed trajectories: $\Delta \varphi =2eAB/\hbar$, where $A$ is the loop area. Here we take $A=\lambda_{\rm SO}^2=2D\tau_{\rm SO}$ as a characteristic ``diffusion area'' probed by our WL/WAL experiment, with $\tau_{\rm SO}$ being the spin dephasing time, and $\lambda_{\rm SO}$ the spin diffusion length. By taking $\Delta \varphi=1$ (rad) at $B = B_{\rm SO}$, we can  extract $\tau_{\rm SO}$ from the minima of the WAL curves, which yields $\tau_{\rm SO} = \hbar (4eDB_{\rm SO})^{-1}$. The factor of 4 here stems from the two time-reversed paths and the diffusion length.

{\bf Effective SO times and lengths.} Theoretically, we determine $\tau_{\rm SO}$ via a spin random walk process (D'yakonov-Perel (DP)). The initial electron spin in a loop can point (with equal probability) along the $s_{x_-}$, $s_{x_+}$, and $s_z$ axes (analogous to $x_+$, $x_-$, and $z$, respectively), which have unequal spin-dephasing times $\tau_{\rm DP,s_{x_-}}$, $\tau_{\rm DP,s_{x_+}}$, and $\tau_{\rm DP,s_z}$. Hence, for unpolarized, independent spins, we take the average $\tau_{\rm eff} =(\tau_{\rm DP,s_{x_-}}+\tau_{\rm DP,s_{x_+}}+\tau_{\rm DP,s_z})/3$ which leads to an average spin dephasing length $\lambda_{\rm eff}= \sqrt{2D \tau_{\rm eff}}$. In the SOM, we discuss the spin random walk and provide expressions for the DP times including corrections due to the cubic $\beta_3$ term. Figure 4 shows curves for the spin dephasing times and lengths presented here. In the main panel, the cubic $\beta_3$ is neglected in $\lambda_{\rm eff}$ since for $n\leq7\,\cdot10^{11}\,\mathrm{cm^{-2}}$, WL appears at $\alpha=\beta$ (small $\beta_3$). In contrast, the cubic term is included in $\tau_{\rm eff}$ in the inset since at the higher density $n=9\,\cdot10^{11}\,\mathrm{cm^{-2}}$, WAL persists (strong $\beta_3$).


\section{Acknowledgments}
We would like to thank A. C. Gossard, D. Loss, D. L. Maslov, G. Salis for valuable inputs and stimulating discussions. This work was supported by the Swiss Nanoscience Institute (SNI), NCCR QSIT, Swiss NSF, ERC starting grant, EU-FP7 SOLID and MICROKELVIN, US NSF and ONR, Brazilian grants FAPESP, CNPq, PRP/USP (Q-NANO), and natural science foundation of China (Grant No.~11004120).

\end{document}